# Krypton oxides under pressure

Patryk Zaleski-Ejgierd [a,*] and Paweł Łata [a,+]

[a] Institute of Physical Chemistry, ul. M. Kasprzaka 44/52, 01-224 Warsaw, Poland

*pze.work@gmail.com, [+]plata@ichf.edu.pl

**ABSTRACT**

*Under high pressure, krypton, one of the most inert elements is predicted to become sufficiently reactive to form a new class of krypton compounds; krypton oxides. Using modern ab-initio evolutionary algorithms in combination with Density Functional Theory, we predict the existence of several thermodynamically stable Kr/O species at elevated pressures. In particular, our calculations indicate that at approx. 300 GPa the monoxide, KrO, should form spontaneously and remain thermo- and dynamically stable with respect to constituent elements and higher oxides. The monoxide is predicted to form non-molecular crystals with short Kr-O contacts, typical for genuine chemical bonds.*

Ever since the successful isolation of argon (1), followed by five other noble-gases, the six members of Group-18 were long considered very much unreactive, if not completely inert. Nonetheless, during the advent of quantum chemistry and the resulting rapid development of theoretical apparatus Pauling did predict possible formation of stable noble-gas compounds, in particular $XeF_6$ and $KrF_6$ (2). It took another 30 years before Bartlett synthesized the first noble-gas compound, $XePtF_6$ (3). The very first synthesis of neutral krypton compound, the difluoride ($KrF_2$), was reported soon-after, in 1963 (4).

In 1992 Lundell and Kanttu studied simple molecular $Kr_2H^+$ using modern *ab-initio* methods, including effective core potential approach and all-electron methods (5). In 2000, Lundell *et al.* investigated experimental formation of more complex, neutral krypton hydrides of the H-Kr-*Y* type (here *Y* is an electro-negative fragment) in low-temperature matrices (6). In 2002 Pettersson *et al.* reported formation of H-Kr-F in low-temperature Kr matrix via VUV photolysis of the HF precursor (7). Later on, in 2003 Räsänen group reported first synthesis of neutral Kr-containing organic compounds, in the form of H-Kr-C≡C-H, followed by an extensive *ab-initio* calculations (8). Most recently Kamenova *et al.* studied in details the mechanism and kinetics of the processes leading to the formation of H-Kr-C≡C-H in solid krypton matrices (9). To date, a number of other krypton species have been identified, including van der Waals complexes (10), numerous salts containing in particular $KrF^+$ and $Kr_2F_3^+$ cations (11) (12) and several complexes containing coordinated neutral F-Kr-F units (13) (14). For more information of krypton chemistry see Nabiev *et al.* (15) and Grochala (16).

Recent developments in the field of ultra-high pressures, in particular development of the diamond-anvil cell (DAC) techniques, allow for synthesis and detailed characterization of new species at unprecedentedly high pressures (17). Nevertheless, while the DAC techniques are becoming increasingly available they are still far from routine and require a significant amount of expertise and experimental effort.

To date, very few krypton compound have been studied at high-pressure (10). To the best of our knowledge no binary compound of krypton and oxygen have been previously investigated. In the absence of experimental data we applied modern computational techniques to explore the static phase diagram of the Kr–$O_2$ system in a wide pressure range, from 10 to 500 GPa, searching for hitherto unknown krypton oxides. We analyzed the possibility of high-pressure formation and stabilization of four binary systems with variable oxygen content, including KrO, $KrO_2$, $KrO_3$ and $KrO_4$. For each stoichiometry we performed an extensive search for the most stable structures combining several complementary approaches, including application of modern evolutionary algorithms based on purely chemical composition (with no experimental input required) (18) (19).

**RESULTS and DISCUSION**

At ambient pressure krypton freezes at 115.77K to form a while crystalline solid with face-centered cubic (*fcc*) crystal structure. In our calculations already at slightly elevated pressure (P>10 GPa) krypton to adopts a hexagonal close-packed (*hcp*) structure and maintains it in the whole considered pressure range (10-500 GPa). In the case of oxygen, depending on the pressure applied, we used ε and ζ structures for reference (20). Concerning the Kr-$O_2$ mixture there are certain trends one might expect. While krypton remains unreactive at ambient conditions a substantial increase of pressure should bring all atoms, both Kr and O closer together, at least on average. Some Kr-O contacts should be more susceptible to compression thus inducing krypton-oxygen reactivity.

To identify thermodynamically stable binary compounds of krypton and oxygen we analyze a number of predefined Kr:O ratios (1:1, 1:2, 1:3 and 1:4) in a wide pressure range (P = 10-500 GPa). We analyze the relative stability of each of the considered stoichiometries by plotting the relative enthalpies of formation, calculated per atom, as a function of oxygen content. The resulting tie-line plot is given in Fig. 1. Oxides which are stable with respect to disproportionation into *other* oxides and/or pure elements form a convex hull of energy with respect to relevant substrates. Note that even at 400 > *P* ≫ 200 GPa all of the studied oxides remain unstable with respect to disproportionation into pure elements.

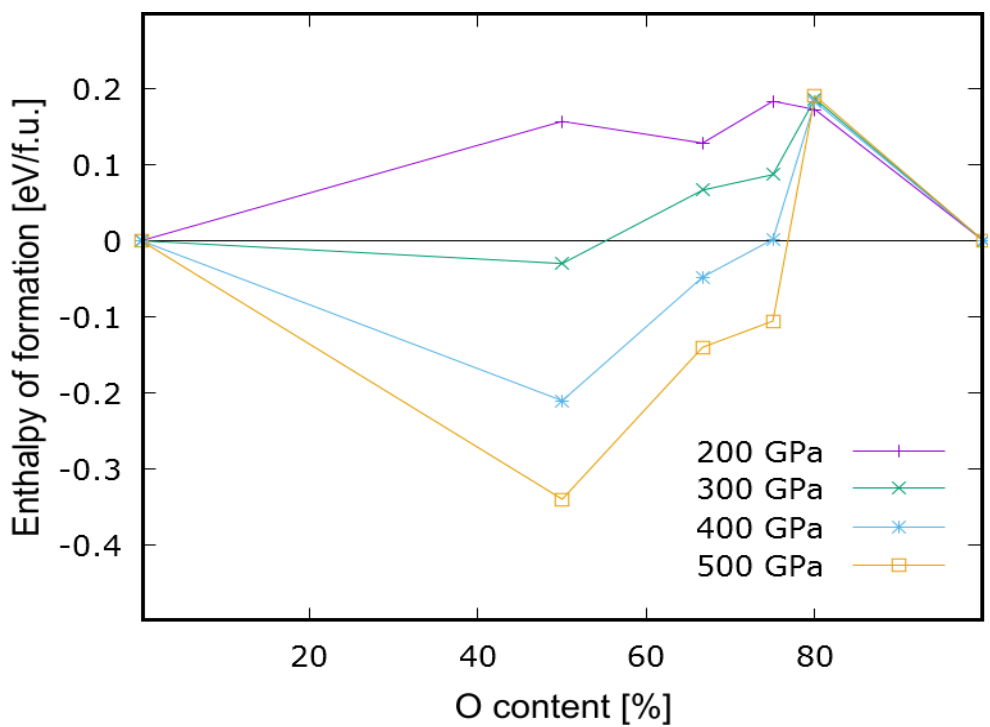

*Figure 1:* *Relative enthalpies of formation (per atom) of the KrO$_x$ phases (x=1-3) calculated with respect to elemental krypton and molecular oxygen in their corresponding most stable phases. Oxides which are stable with respect to disproportionation are those forming a convex hull of enthalpy with respect to composition into selected species. Note the substantial stabilization of the monoxide.*

The *mono*xide is the first to stabilize, at P ≈ 285 GPa, and it remains the primary candidate for synthesis as it is the only stoichiometry thermodynamically stable not only with respect to pure elements but *also* with respect to disproportionation into other oxides. As the pressure increases, the relative enthalpy of formation of higher oxides, with the notable exception of KrO$_4$, become less and less positive and eventually negative but it still remains substantially negative only for KrO. At the level of theory applied our calculations clearly indicate that at a sufficiently high pressure krypton oxide formation should be thermodynamically viable with the 1:1 stoichiometry clearly favored. With this in mind we now proceed with the detailed characterization of the predicted, most stable phases of KrO. In Fig. 2 we present the relative enthalpy of formation calculated for the four most competitive phases of KrO.

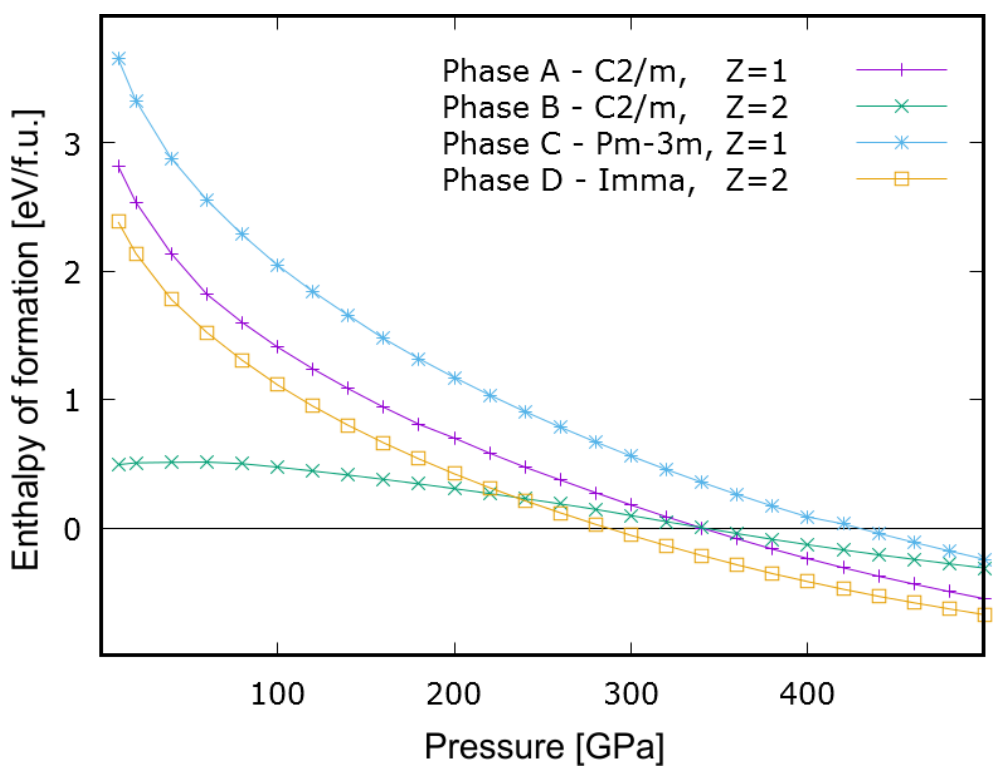

*Figure 2: The relative enthalpies of formation, ΔH [eV/f.u.], calculated for selected phases of KrO with respect to pure elements in their thermodynamically most stable states. For Phase D, the theoretically calculated stabilization pressure is approx. 285 GPa (for details on the structures see Fig. 3 and Table S1).*

As the pressure increases, the relative enthalpies of formation of all the phases become negative. Phase A and D exhibit non-molecular character, Phase B is a mixture of atomic krypton and molecular oxygen and Phase C is a host-guest structure (see Fig. 3 for schematic illustration). The non-molecular Phase D stabilizes first, at approximately 285 GPa, and remains the most stable KrO phase up to 500 GPa.

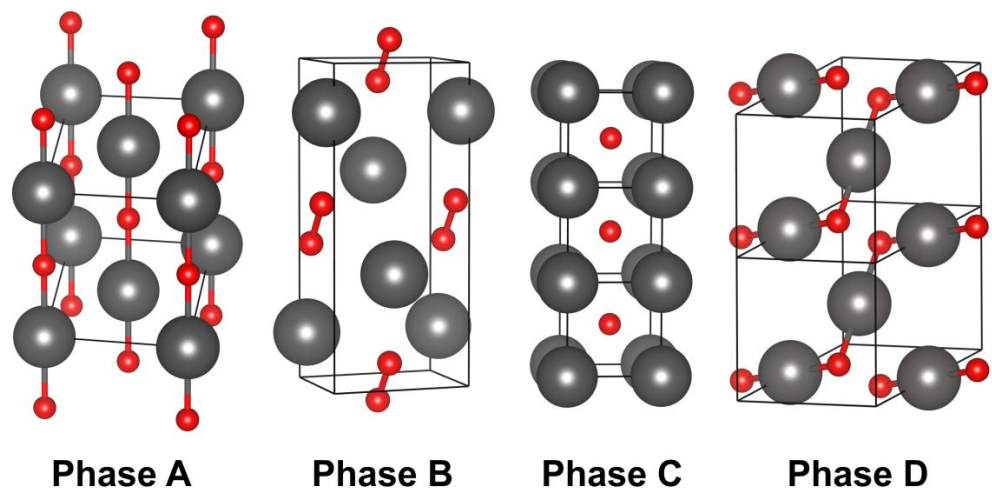

*Figure 3: Schematic representation of the selected most-stable phases of KrO and their corresponding unit cells (large gray spheres – krypton; small orange spheres – oxygen). For illustration purposes, and clarity, selected Kr-O and O-O contacts are depicted (Kr-O < 2.0 Å and O-O < 1.6 Å). Note the significantly different character of the given phases. For details on bonding see Table S1; for details on the stability see Figure 2.*

At low pressures Phase B is thermodynamically most stable although its relative enthalpy of formation with respect to bulk krypton and oxygen remains positive up to 340 GPa. This metastable, molecular phase consists of atomic krypton mixed with *diatomic* oxygen. Even at 300 GPa the oxygen molecule is clearly present with the O-O bond length equal to 1.235 Å (compare to 1.210 Å at ambient pressure). At the same time the shortest Kr-O contact distance is equal to 2.119 Å, indicating lack of significant chemical interaction.

At 285 GPa Phase D stabilizes. While it remains highly unstable at low pressure, upon stabilization it remains the most stable even at 500 GPa. Phase D lacks the molecular character. At 300 GPa the shortest O-O distance is 2.244 Å while the shortest Kr-O contact is only 1.807 Å indicating significant interaction between oxygen and the noble gas atoms. Structurally, Phase D consists of infinite ⟍Kr ⁄O⟍Kr ⁄O⟍ zig-zag chains with the ∠(Kr-O-Kr)=84.35° (see Fig. 3 and Fig. 6 for illustration).

Phase A shares several similarities with phase D; it also lacks any molecular character and it is also constructed from infinite, albeit linear, -Kr-O-Kr-O- chains; within each such chain all Kr-O distances are equal. At 300 GPa the shortest O-O distance is now 2.455 Å while the shortest, intrachain Kr-O contacts are 1.863 Å.

Phase C is given for completeness. In contrast to the other phases it is as a host-guest-type structure in which each krypton atom is coordinated by eight oxygen atoms. At 300 GPa the eight Kr-O distances are 2.126 Å long and are comparable to Kr-$O_2$ distance (2.119 Å) in the molecular Phase B – once again indicating of lack of substantial Kr-O interactions.

Among the four discussed phases the shortest observed Kr-O distance (1.807 Å) is present in Phase D. Only a slightly longer distance is also present in Phase A (1.863 A). In both the phases the second shortest Kr-O contact is 2.185 and 2.126 Å respectively and it is nearly identical with the shortest Kr-O contact in Phase B and C (2.119 and 2.126 Å respectively) in which oxygen, either molecular or atomic, is expected to be chemically unbound.

Here we note that the shortest Kr-O contact in Phase D (1.807 Å) is only 13.6% longer than a single covalent bond estimated by Pyykkö and Atsumi (1.590 Å) (21). The presence of such short directional contacts indicates genuine chemical bonding between krypton and oxygen and explains increased theromodynamical stability of Phase A and D.

Indeed, among the best structures for higher krypton oxides ($KrO_x$, x=2, 3, 4) we observed exclusively *molecular* crystals in which monoatomic krypton is mixed with molecular oxygen (see Fig. S1 for detail). For those oxygen-rich stoichiometries we find no evidence of short Kr-O bonding, even at 500 GPa.

Phonon calculations are of interest both as guarantors of dynamical stability and with respect to the information such calculations may yield on the ease, or difficulty, of motions in the structures considered. Our analysis reveals that all of the reported KrO phases are dynamically stable within the pressure range corresponding to their theromodynamical stability. For all the considered phases, except Phase C, we observed negative frequencies only at P<<250 GPa. Phase C does stabilize dynamically but only at P>>400 GPa; we thus neglect it in the further considerations. In Fig. 4 we present phonon analysis for Phases A, B, and D performed consistently at P=400 GPa (sufficiently high for all the phases to become thermodynamically stable).

For Phase A we observe no separate phonon branches. Visualization reveals numerous modes in which oxygen and krypton atom motion is always conjugated. Interestingly, we observe one particularly flat phonon band giving rise to sharp increase in the phonon density of states (PhDOS) at 600 $cm^{-1}$. We identify this contribution as a mixture of stretching and bending modes of -O-Kr–O-. In case of Phase B a clearly separated phonon branch is located at ca. 1550 $cm^{-1}$ corresponding to the stretching mode of the $O_2$ molecules. For Phase D we observe two separate branches; the high energy branch (700-1000 $cm^{-1}$) arises primarily due to oxygen contributions while the low energy branch (0-600 $cm^{-1}$) arises due to krypton displacements.

We note that the higher oxides, in particular $KrO_2$, are also dynamically stable upon theromodynamical stabilization. In case of the enthalpically best structures (see SM, Fig. S1) our vibrational analysis reveals presence of the stretching modes located at approx. 1550 $cm^{-1}$ clearly indicating presence of the undissociated $O_2$ molecules.

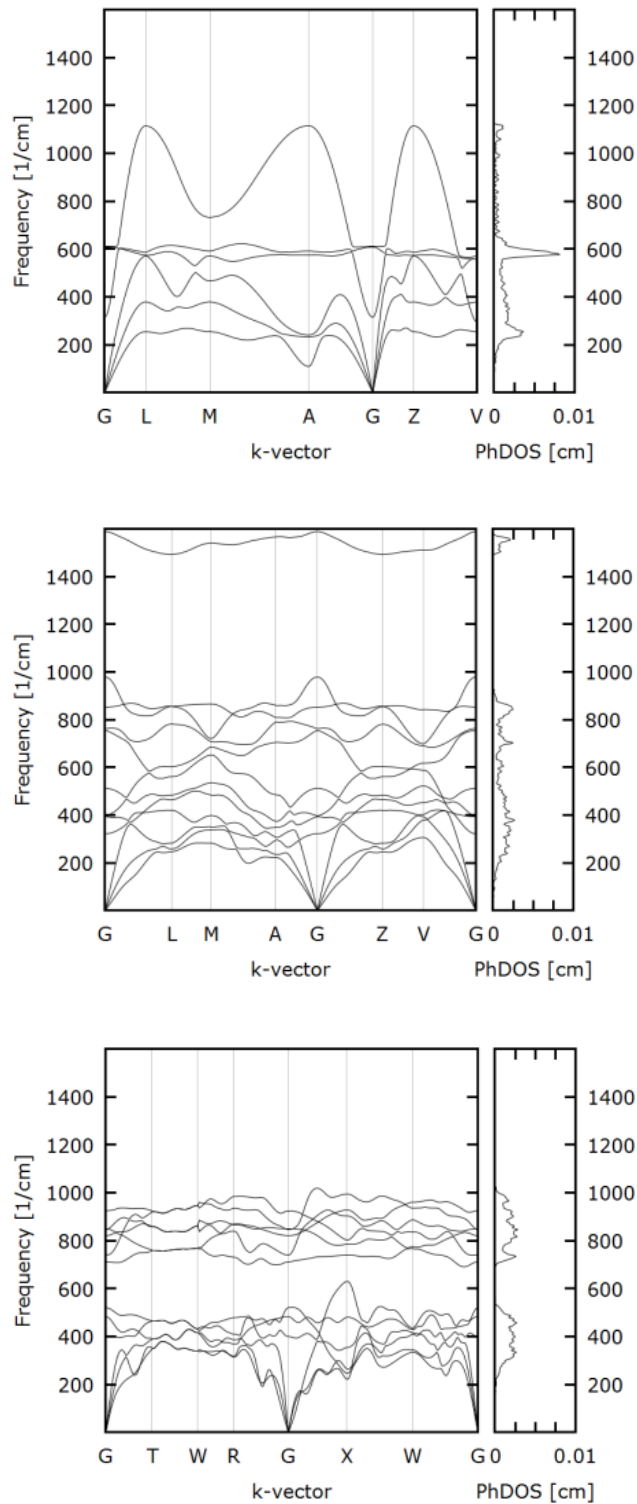

*Figure 4:* Phonon dispersion and the corresponding phonon density of states (PhDOS) calculated at 300 GPa for selected structures: Phase A (C2/m, Z = 1) – top panel, Phase B (C2/m, Z = 2) – middle panel and Phase D (Immm, Z = 2) – bottom panel. Note the presence of the O-O stretching band due to molecular oxygen (middle panel).

We analyze the conductive properties of the identified phases of KrO by calculating band structures and the corresponding electronic densities of states (DOS) for the ground state structures optimized at 300 GPa. Here, we note, that the Kohn–Sham formulation of DFT is well known to underestimate the width of electronic band gaps. While we did not try to correct for this, the study of the electronic structures is still likely to give us reliable, qualitative information about the insulating/semiconducting/metallic character of the considered krypton oxide phases. We here focus on the characterization of the three viable structures for 1:1 stoichiometry: Phase A, B and D.

The results are shown in Fig. 5. In all three cases analysis of the total DOS indicates significant accumulation of the occupied states below the Fermi level ($E_f$); only for Phase A the DOS remains non-zero at Fermi level and is indicative of metallic character. This metallic character is likely a derivative of the polymeric structure of Phase A composed of infinite linear -(Kr-O)$_n$- chains. For the two other Phases, B and D, electronic gaps are observed, indicating that the two phases are most likely either insulating or, particularly in the case of Phase B, semi-conductive. In all three cases the valence band levels compose primarily from the *p*-orbitals of krypton and oxygen. The partitioning of the total densities of states along with the Mullikan and Hirschfield analysis suggest significant charge transfer from krypton to oxygen atoms.

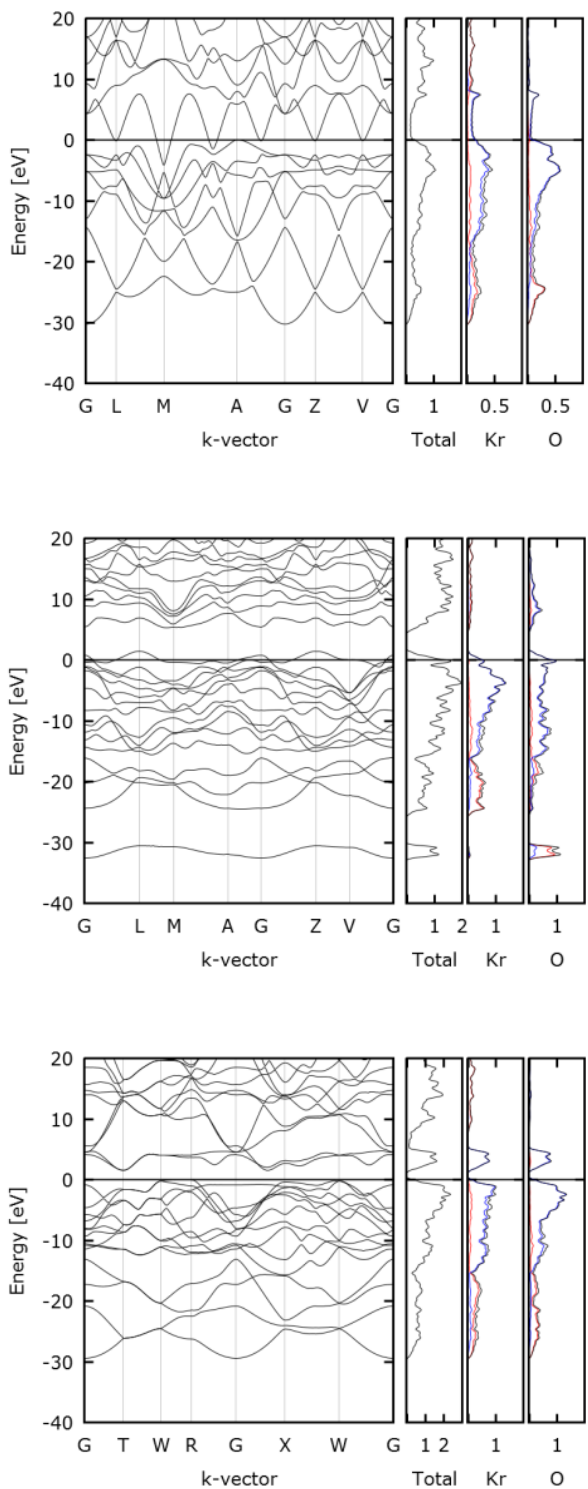

*Figure 5*: Band structures, total density (DOS) and the partial densities of states (PDOS) [el./eV] calculated for: Phase A (C2/m, Z=1) – left panel, Phase B (C2/m, Z=2) – middle panel and Phase D (Immm, Z=2) – right panel (red: s - character, blue: p - character, black: total DOS). The Fermi energy ($E_f$) is set to zero for convenience.

Altogether the results indicate that under pressure krypton loses its chemical inertness. In both the non-molecular phases (Phase A and D) krypton is expected to form genuine bonds with oxygen. To complete the description of the predicted oxides and to illustrate bonding in Kr-O we plot the electron density distribution function. For consistency we chose the same pressure and the same cut-off values for all three cases (300 GPa and 0.15 $a_0^{-3}$, respectively). The results are shown in Fig. 6.

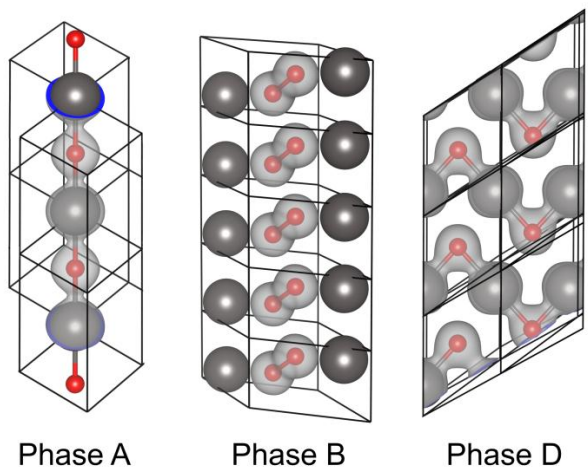

*Figure 6: The three enthalpically best structures calculated for KrO at 300 GPa along with the superimposed electron density isosurfaces (we set the cut-off value consistently at 0.15 $a_0^{-3}$, $a_0$ = Bohr radius). The short Kr-O bonds (R < 1.9 A) and the O-O bonds are also depicted; note the increased electron density distribution precisely along the bonds.*

Note the presence of significantly increased electron density between krypton and oxygen atoms. Similarly, in case of Phase B (the molecular one) electron density distribution provides clear evidence on the interstitial character of the guest oxygen molecules.

In summary, we predict that at high compression (P=285 GPa) krypton monoxide stabilizes thermodynamically and forms spontaneously. At even higher pressures (P approx. 400 GPa) a di- and trioxide also stabilize. The monoxide is expected to remain the preferred stoichiometry both with respect to decomposition into pure elements and with respect to disproportionation into the higher oxides even at the highest of the considered pressures (up to 500 GPa). The most stable phase predicted for krypton monoxide is a non-molecular Phase D (Imma, Z=2); in it krypton forms genuine chemical bonds with oxygen. At 300 GPa our calculations identify short Kr-O contacts in the order of 1.80 Å; within the range expected for covalent bonds.

We note that this prediction can be tested experimentally; in particular, within all the predicted stoichiometries and the corresponding phases, structure of the krypton framework is a clear departure from the high-pressure *hcp* structure of bulk krypton; should one try to synthesize the oxides, such dramatic symmetry change should help identify the resulting species.

**ACKNOWLEDGEMENTS**

The grant UMO-2012/05/B/ST3/02467 of the Polish National Science Centre (NCN) is acknowledged.

**CONTRIBUTIONS**

The manuscript was written through contributions of all authors. All authors have given approval to the final version of the manuscript. P.Z.-E. conceived the research. P. Ł. and P.Z.-E. performed relevant DFT calculations. P.Z.-E. wrote the article.

**COMPETING FINANCIAL INTERESTS**

The authors declare no competing financial interests.

**CORRESPONDING AUTHORS**

Patryk Zaleski-Ejgierd (pzaleski@ichf.edu.pl), Paweł Łata (plata@ichf.edu.pl)

**METHODS**

All of the reported calculations were performed using density functional theory (DFT) with the Perdew–Burke–Ernzerhof (PBE) parameterizations of the generalized gradient approximation (GGA) as implemented in the VASP code (ver. 4.6) (22) (23). The projector augmented wave (PAW) (24) (25) method was applied with PAW pseudo-potentials taken from the VASP archive. For the plane-wave basis-set expansion, an energy cut-off of 1000 eV was used. Valence electrons (Kr: $4s^24p^6$ ; O: $2s^22p^4$) have been treated explicitly, while standard VASP pseudopotentials were used for the description of core electrons.

The Γ-centered k-point mesh was generated for every structure in such a way that the spacing between the k points was approximately 0.10 Å. Structures have been optimized using a conjugate-gradient algorithm with a convergence criterion of $10^{-7}$ eV. This ensured that in the optimized structures forces acting on the ions were in the order of 1 meV·Å$^{-1}$. The enthalpies derived from our calculations do not contain the zero-point-energy (ZPE) corrections. Contrary to the hydrogen-rich species, we expect those to be of little importance. For each stoichiometry we performed an extensive computational search for the most stable structures combining several complementary approaches: prototypical structures, purely random structure generation and evolutionary algorithms (18) (19). In our calculations we used up to Z=4 formula units per unit cell.

Dynamical stability of the identified, enthalpically preferred structures has been assessed through phonon analysis within the harmonic approximation. Phonon dispersion curves were calculated using the finite-displacement method as implemented in the CASTEP code . We used sufficiently large super-cells, typically in the order of 3×3×3, or larger, in the interpolation of the force constants required for the accurate phonon calculations. Finally, we stress that the structures we consider are all ground state arrangements whereas the measurements are carried out at finite, typically close to room, temperatures.